\documentclass[prl,twocolumn,showpacs,amsmath,amssymb,superscriptaddress]{revtex4} 

\usepackage{graphicx}    
\usepackage{dcolumn}     
\usepackage{bm}          
\usepackage{subfigure}   
\usepackage[normalem]{ulem}
\usepackage{xcolor}

\newcommand{\hide} [1] { }                                      

\graphicspath{{Figures/},.}
\newcommand{\vwi}{Dresden University of Technology, Andreas-Schubert-Stra{\ss}e\ 23,
                        D-01086 Dresden, GERMANY}
\newcommand{\Bremen}{Institute for Theoretical Physics,
                     University of Bremen, 
                     Otto Hahn Allee 1,
                     D-28334 Bremen,
                     GERMANY }
\newcommand{\NTNU}{Department of Physics,
    Norwegian University of Science and Technology~(NTNU),
     NO-7491 Trondheim, NORWAY}
\newcommand{\uz}{University of Zilina, Univerzitna 8215/5, SK-01026 Zilina, SLOVAKIA}

\begin{document}


\title{Dynamic Effects Increasing Network Vulnerability to Cascading Failures}

\author{Ingve Simonsen}
\email{Ingve.Simonsen@phys.ntnu.no}
\affiliation{\vwi}
\affiliation{\NTNU}

\author{Lubos Buzna}
\affiliation{\vwi}
\affiliation{\uz}

\author{Karsten Peters}
\affiliation{\vwi}

\author{Stefan Bornholdt}
\affiliation{\Bremen}

\author{Dirk Helbing}
\affiliation{\vwi}

\begin{abstract}
  We study cascading failures in networks using a dynamical flow model
  based on simple conservation and distribution laws to investigate
  the impact of transient dynamics caused by the rebalancing of loads
  after an initial network failure (triggering event). It is found
  that considering the flow dynamics may imply reduced network
  robustness compared to previous static overload failure models.
  This is due to the transient oscillations or overshooting in the
  loads, when the flow dynamics adjusts to the new (remaining) network
  structure.  We obtain {\em upper} and {\em lower} limits to network
  robustness, and it is shown that {\it two} time scales $\tau$ and
  $\tau_0$, defined by the network dynamics, are important to consider
  prior to accurately addressing network robustness or vulnerability.
  The robustness of networks showing cascading failures is generally
  determined by a complex interplay between the network topology and
  flow dynamics, where the ratio $\chi=\tau/\tau_0$ determines the
  relative role of the two of them.
 \end{abstract}

\date{\today}

   
\maketitle


Societies rely on the stable operation and high performance of complex
infrastructure networks, which are critical for their optimal
functioning. Examples are electrical power grids, telecommunication
networks, water, gas and oil distribution pipelines, or road, railway
and airline transportation networks. Their failure can have serious
economic and social consequences, as various large-scale blackouts and
other incidents all over the world have recently shown.  It is
therefore a key question how to better protect such critical systems
against failures and random or deliberate attacks
\cite{Book:Kaufmann-1977,Barabasi-2002,Newman-2003}.  Issues of
network robustness and vulnerability have not only been addressed by
engineers \cite{Book:Billinton-1994, Dai-2001}, but also by the
physics community~\cite{Barabasi-2002,Newman-2003,Albert-2000,Watts-2002,Holme-2002,Motter-2002,Motter-2004,Albert-2004,Crucitti-2004,Scire-2005,Huang-2006,Bakke-2006,DallAsta-2006,Kaluza-2007}.
In the initial studies of this
kind~\cite{Barabasi-2002,Albert-2000,Holme-2002,Watts-2002,Albert-2004},
the primary concern was dedicated to what can be termed {\it
  structural robustness}; the study of different classes of network
topologies and how they were affected by the removal of a finite
number of links and/or nodes ({\it e.g.} how the average network
diameter changed).  It was concluded that the more heterogeneous a
network is in terms of, {\it e.g.}, degree distribution, the more
robust it is to random failures, while, at the same time, it appears
more vulnerable to deliberate attacks on highly connected
nodes~\cite{Albert-2004,Watts-2002}.

Later on, the concepts of network loads, capacities, and overload
failures were
introduced~\cite{Motter-2002,Motter-2004,Crucitti-2004,Huang-2006,Bakke-2006,DallAsta-2006}.
For networks supporting the flow of a physical quantity, the removal
of a node/link will cause the flow to redistribute with the risk that
some other nodes/links may be overloaded and failure prone.  Hence a
triggering event can cause a whole sequence of failures due to
overload, and may even threaten the global stability of the network.
Such behavior has been termed {\it cascading failures}. A seminal work
in this respect is the paper by Motter and Lai~\cite{Motter-2002}.
These authors defined the load of a node by its betweenness
centrality~\cite{Motter-2002,Newman-2003}.  Subsequent studies
introduced alternative measures for the network
loads~\cite{Crucitti-2004} as well as more realistic redistribution
mechanisms~\cite{Crucitti-2004,Bakke-2006,Huang-2006,DallAsta-2006}.

In all studies cited above, the redistribution of loads is treated
time-independent or static. We will refer to them collectively as {\it
  static overload failure models}.  The load redistributions in such
models are instantaneously and discontinuously switched to the
stationary loads of the new (perturbed) network, {\it i.e.}  the {\it
  transient} dynamical adjustment towards the new stationary loads of
the perturbed network is neglected.

\begin{figure}
  \begin{center}
    \includegraphics*[width=0.95\columnwidth, height=0.6\columnwidth]{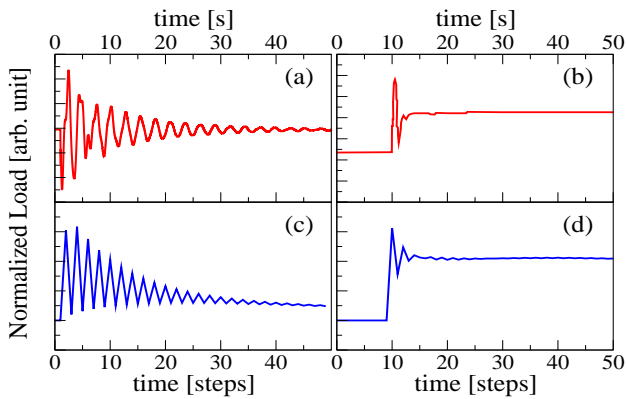}
  \end{center}
  \caption{\label{Fig:compare} (Color online) Comparison of the
    time-dependent link loads after an triggering event (taking place
    at $t=0$) as predicted by state-of-the-art power simulators
    (Figs.~\protect\ref{Fig:compare}(a)~\cite{Sadikovic-2006} and
    (b)~\cite{eurostag}), and the simple ``flow-conserving'' model
    described and used in the present work
    (Figs.~\protect\ref{Fig:compare}(c) and (d)).}
\end{figure}

The aim of this Letter is to compare robustness estimates of complex
networks against cascading failures where the dynamical flow
properties are taken into account relative to those where they are not
(static case).  This work does {\em not} intend to target a specific
system (or network); instead we aim at being as generic as possible in
the choice of dynamical model with the consequence that particular
details and features of a specific system have to be neglected, {\it
  i.e.}\ we work with a minimal model as often favored in physics.
Nevertheless, the conceptually simple dynamical phenomenological flow
model that we propose, incorporates {\em flow conservation}, {\em
  network topology}, as well as {\em load redistribution} features
that are shared by real-life systems. On this background, it is
expected ({\it cf.} Fig.~\ref{Fig:compare}) that the model results
will reflect some important properties of real-life systems.
  
For matters of illustration and to facilitate comparison with previous
results~\cite{Motter-2002,Motter-2004,Crucitti-2004,Huang-2006,Bakke-2006,DallAsta-2006},
we have worked with topologies of power transmission networks.
Although our model seems to capture stylized features of electrical
networks (see Fig. \ref{Fig:compare}), we stress that our goal is {\it
  not} a realistic representation of those, nor is our model
restricted to such systems.  Within the proposed model, we want to
demonstrate that time-dependent adjustments can play a crucial role.
In particularly, we will show that static overload failure models give
the {\it lower} limit of the vulnerability of flow networks to
failures and attacks, and hence of the probability of cascading
failures.

\par
In order to study this, in the very tradition of physics, we use a
simple flow model with few parameters, which however considers the
network topology, flow conservation, and the distribution of loads
over the neighboring links of a node
\cite{Eriksen-2003,Simonsen-2005}.  We assume a network consisting of
${\cal N}$ nodes and represent it by a matrix $\boldsymbol{W}$, whose
entries $W_{ij}\geq 0$ (with $i,j=1,2,\ldots,{\cal N}$) shall reflect
the weight of the (directed) link {\it from} node $j$ to $i$ (with
$W_{ij}=0$ indicating no link present). The relative weights
$T_{ij}=W_{ij}/ w_i$ shall define the elements of the transfer matrix
$\boldsymbol{T}$, where $w_j=\sum_{i=1}^{{\cal N}} W_{ij}$ is the
total {\em outgoing} weight of node $j$~\cite{Simonsen-2005}. These
elements describe the distribution of the overall flow (per unit
weight) $c_j(t)$ reaching node $j$ at time $t$ over the neighboring
links $i$.  When the flow is assumed to reach the neighboring nodes
$i$ at time step $t+1$, we obtain $c_i(t+1) = \sum_{j=1}^{{\cal N}}
T_{ij} c_j(t) + j_i^\pm$~
\cite{Eriksen-2003,Simonsen-2005,Helbing-1996}, where we have added
possible source terms ($j_i^\pm >0$) or sink terms ($j_i^\pm < 0$). In
vectorial notation, the network flow equation reads
\begin{align}
  \label{eq:MarkovChain-current}
  \boldsymbol{c}(t+1) &= \boldsymbol{T} \boldsymbol{c}(t) +
  \boldsymbol{j}^\pm \, ,
\end{align}
resembling Kirchhoff's first law from circuit theory.

\par
\begin{figure}
  \begin{center}
    \includegraphics*[width=0.4\columnwidth]{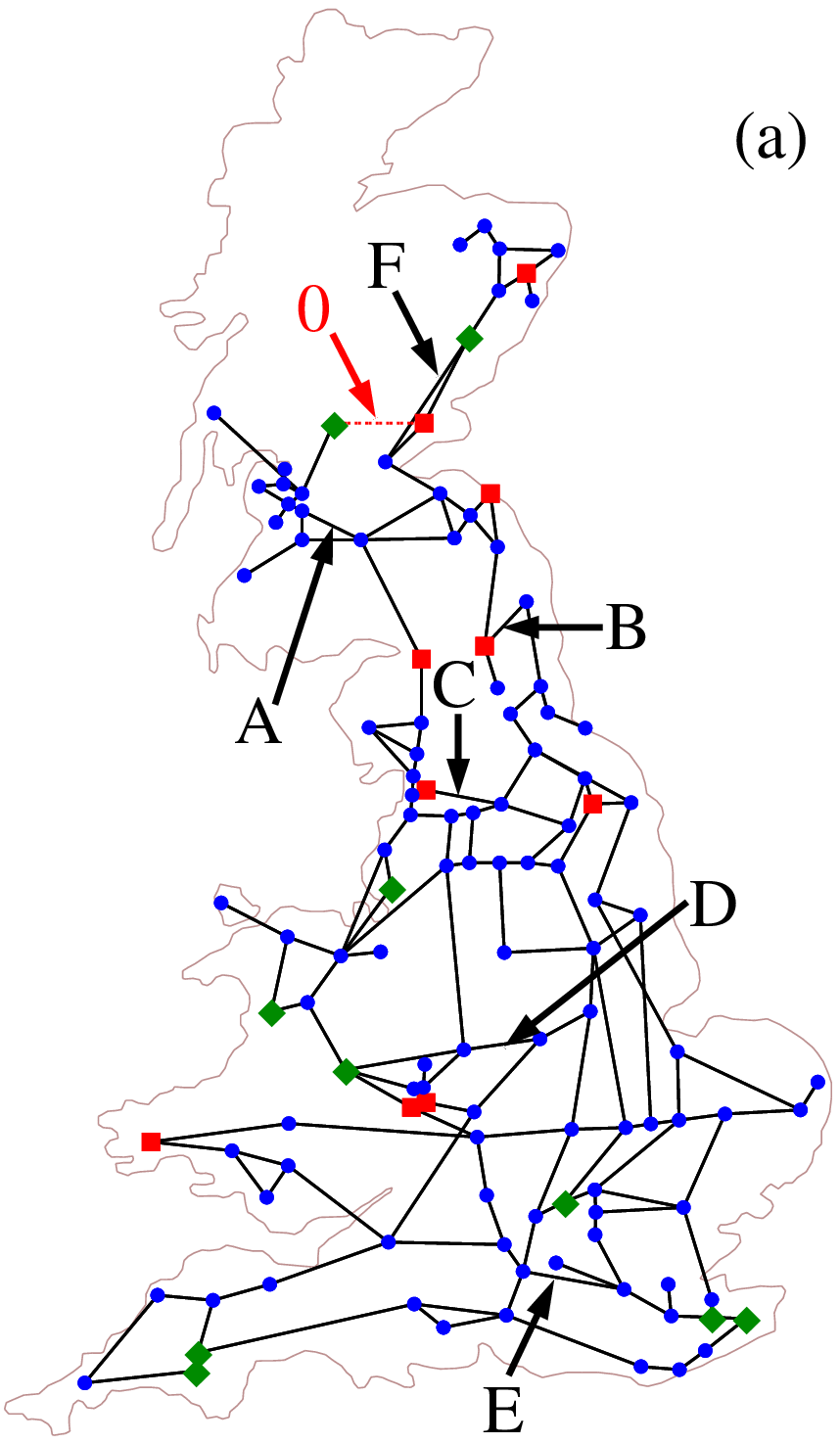}\\*[0.1mm]
    \includegraphics*[width=0.9\columnwidth, height=0.35\columnwidth]{Fig2b}\\*[-0.1mm]
    \includegraphics*[width=0.9\columnwidth, height=0.35\columnwidth]{Fig2c}
  \end{center}
  \caption{\label{Fig:UK} (Color online) (a) Illustration of the
    dynamics of our network flow model assuming the topology of the UK
    high-voltage power transmission grid ($300$--$400$~kV) consisting
    of $120$ geographically correctly placed nodes (generators,
    utilities, and transmission stations) and $165$
    links~(transmission lines).  The network was treated as unweighted
    and undirected. In our simulations, twenty of the existing network
    nodes were chosen randomly to play the role of generator~(source)
    and utility~(sink) nodes ($\left|n_i^\pm/{\cal
        N}\right|=2.5\cdot10^{-4}$), ten of each kind. In
    Fig.~\protect\ref{Fig:UK}(a) the location of these nodes are
    indicated by filled red squares and filled green diamonds
    respectively.  At time $t=0$, before which the network loads were
    in the stationary state ($L_{ij}(t<0)$), the network was perturbed
    by removing a transmission line in Scotland (the red dashed link
    marked by 0 in Fig.\protect\ref{Fig:UK}(a)).  The resulting
    normalized transient link loads, $L_{ij}(t)/L_{ij}(t<0)$, are
    depicted in Figs.~\ref{Fig:UK}(b) and (c) for some selected links
    of the UK transmission grid, as indicated in
    Fig.~\protect\ref{Fig:UK}(a).  The horizontal dash-dotted lines
    correspond to the normalized stationary loads of the links.  }
\end{figure}

If $\boldsymbol{j}^\pm=0$, the stationary solution to
Eq.~\eqref{eq:MarkovChain-current} is a constant vector with
components $c_i^{(0)}(\infty)\sim 1/\sqrt{{\cal N}}$, while with a
source term present ($\boldsymbol{j}^\pm \neq 0$), it can be expressed
as $\boldsymbol{c}(\infty)= \boldsymbol{c}^{(0)}(\infty)+
\left(\boldsymbol{1}-\boldsymbol{T}\right)^+\boldsymbol{j}^\pm$ with
$\left(\boldsymbol{1}-\boldsymbol{T}\right)^+$ denoting the so-called
generalized inverse~\cite{Book:PsudoInverse} of the singular matrix
$\boldsymbol{1}-\boldsymbol{T}$.  Hence, the total directed current on
link $j\rightarrow i$ at time $t$ becomes $C_{ij}(t)=W_{ij}c_j(t)$,
from which also the (undirected) {\it load} $L_{ij}(t)$ of this link
can be defined {\it via} $L_{ij}(t) =
C_{ij}(t)+C_{ji}(t)$~\footnote{Alternatively one could have defined
  directed loads by $L_{ij}(t) = C_{ij}(t)$, but this possibility will
  not be considered herein.}. Closed-form expressions for the flow
dynamics at single nodes have been derived in
Ref.~\cite{Helbing-1996}.  These allow one to study the wave-like
spreading and dissipation of perturbations in the network while
propagating via neighboring links, second-next, {\it etc.} (see Fig.
\ref{Fig:UK}).  Such perturbations may result from the redistribution
of flows after the failure of an overloaded link.

\par
In the seminal work of Motter and Lai~\cite{Motter-2002}, failure of a
node was based on the long-term overload, {\it i.e.} a node was assumed to
fail whenever the {\it stationary} load in the perturbed network
(considering previously broken nodes) exceeded the node capacity. The
(node) capacities were defined as $1+\alpha$ times the (stationary)
loads of the {\it original} network with $\alpha\geq0$ being a global
{\it tolerance factor} ({\it i.e.} a relative excess capacity or safety
margin). In other words, the evaluation of overloading was previously
done {\it after} the system relaxed, without considering the
time-history of how it got to this state ({\it static overload failure
  models})~\cite{Motter-2002,Motter-2004,Crucitti-2004,Bakke-2006,Huang-2006,DallAsta-2006}.

\par
In this Letter, we generalize this approach towards a {\it dynamical
  overload failure model}.  Specifically, in our computer simulations
we assumed a link from node $j$ to $i$ to be overloaded (and to fail)
whenever the time-dependent load $L_{ij}(t)$ exceeded the link
capacity $\mathcal{C}_{ij}$ for at least a time period $\tau$, the
{\it overload exposure time}. The {\it link} capacities were defined
analogously to Motter and Lai \cite{Motter-2002} as
\begin{align}
  \label{Eq:Capacities}
  \mathcal{C}_{ij} &= \left(1+\alpha\right)L_{ij},
\end{align}
where $L_{ij}$ denote the stationary loads of the original network.

\par
In the following, we study the transient dynamical effects and
overload situations that may occur {\it before} the stationary state
is reached.  While for $\tau=0$, a failure results immediately after a
first-time overload, $\tau>0$ implies that the system will have to be
overloaded for a certain time period in order to cause a failure. The
static overload failure model corresponds to $\tau\rightarrow \infty$,
or in practice, $\tau \gg \tau_0$, where $\tau_0$ denotes the
transient time of the system (the inverse of the smallest non-zero
eigenvalue of $\boldsymbol{T}$).  Therefore, the ratio
$\chi=\tau/\tau_0$ can be used to interpolate between the static
($\chi\rightarrow\infty$) and (instantaneous) dynamical overload
failure ($\chi=0$) models. While the static overload failure model
describes the {\it upper} limit of network robustness (the {\it best
  case}), the dynamic overload failure model with $\tau = 0$
gives the {\it lower} limit (the {\em worst case}) due to an 
over\-shooting flow dynamics (Fig.~\ref{Fig:UK}). Realistic cases
are expected to lie between these two limiting cases, corresponding to
a finite value of $\chi$.
  
\par
Apart from network robustness to overload failures, the value of
$\chi$ also determines the {\it dynamics} of failure cascades.  In the
dynamic case with $\chi = 0$, close-by links are more likely to
be overloaded and to fail than in the static case ($\chi \rightarrow
\infty$). Therefore, in the dynamic scenario one tends to have a
pronounced ``failure wave'' sweeping over the network.

\begin{figure}
 \begin{center}
   \includegraphics*[width=0.90 \columnwidth, height=0.50\columnwidth]{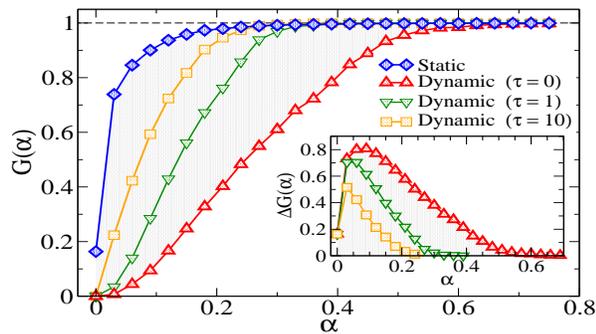}
   \caption{\label{Fig:US} (Color online) The robustness of the
     Northwestern US power transmission grid~\cite{Watts-1998},
     consisting of $4941$ nodes, with an average node degree of $2.67$
     ({\it cf.}  also
     Refs.~\cite{Amaral-2000,Albert-2004,Bakke-2006}).  The average
     fraction of links (or nodes), $G(\alpha)$, remaining in the giant
     component of this network (after cascading) is depicted as
     function of the tolerance parameter $\alpha$, using the {\it
       static} and {\it dynamic} overload failure models described in
     the text. To obtain these results, the links were assigned
     weights, drawn from a uniform distribution on the interval
     $[1,10]$, and $200$ generator and utility nodes of strength
     $\left|n_i^\pm/{\cal N}\right|=10^{-8}$ were assigned randomly
     ($100$ of each type).  The results were obtained by averaging
     over all possible triggering events (single link removals).  The
     inset shows the difference, $\Delta G(\alpha)$, between the
     static and dynamic overload failure models.  }
 \end{center}
\end{figure}

In order to further illustrate the difference between the static and
dynamic cascading failure models, as well as getting a quantitative
measure of the level of overestimation of robustness, we investigate
one of the networks already studied by Motter and
Lai~\cite{Motter-2002} --- the Northwestern American power
transmission network obtained from Ref.~\cite{Watts-1998}~(see also
Refs.~\cite{Amaral-2000,Albert-2004}).  To evaluate the effect of an
initial network perturbation and the following cascade (if any), we
study the fraction of nodes and links, $G_N(\alpha)$ and
$G_L(\alpha)$, respectively, remaining in the giant component of the
network after potential cascading failures have ceased, which have
been initiated by the random failure of a link~\cite{Motter-2002}.
Both quantities behave 
similarly~\footnote{This is a consequence of the network being either
  almost unaffected by an initial link removal, or experiencing global
  failure where the whole network collapses.}. They are displayed in
Fig.~\ref{Fig:US} for the US power transmission network as functions
of the tolerance parameter $\alpha$.  It has been checked and found
that our static overload failure model well reproduces the general behavior
previously reported in Ref.~\cite{Motter-2002}. Namely, global
cascading failure will occur under random attacks (or failures) mainly
for {\it heterogeneous} networks.  

According to Fig.~\ref{Fig:US}, there is a pronounced difference
between the static and (instantaneous) dynamic overload failure model,
corresponding to upper and lower estimates to the network robustness.
As is shown in the inset to Fig.~\ref{Fig:US}, it can be as
significant as $80\%$, and for more homogeneous link weights we have
found differences even higher than $95\%$. Only for quite significant
tolerance factors ($\alpha\geq 50\%$), the discrepancy between the two
estimates becomes insignificant.  Moreover, it has also been found
that the static model tends to be more sensitive to the location of
sources (and sinks).  Thus, our results show that the role of the
dynamical process taking place on the network can be important when
estimating the robustness of networks to failures and random attacks.
It is not only the topology of the network that matters, but also the
properties of the network {\it dynamics} as measured by $\chi =
\tau/\tau_0$.  The change in one or both of them will require a new
robustness estimate.

\smallskip 

In conclusion, we have simulated a simple network flow model
considering, besides network topology, a flow-conserving dynamics and
distribution of loads.  Within this framework, we have studied the
role of the transient dynamics of the redistribution of loads towards
the steady state after the failure of network links. This transient
dynamics is often characterized by overshootings and/or oscillations
in the loads, which may result in characteristic ``failure waves''
spreading over the network. We have furthermore found that,
considering only the loads in the steady state (the static overload
model), gives a best case estimate (upper limit) of the robustness.
The worst case (lower limit) of robustness can be determined by the
instantaneous dynamic overload failure model and may differ
considerably.

Our simple dynamical approach provides additional insights into
systems in which network topology is combined with flow, conservation
and distribution laws. These are potentially useful to understand,
better design and protect critical infrastructures against failures.
For instance, overloads related to high electrical currents cause
(through over-heating of wires) a slow spreading of failures as
compared to the adjustment dynamics of the currents.  This corresponds
to $\chi \gg 1$. In contrast, within the validity limits of Ohm's law,
one may also use our model to mimic effects of overloads related to
overvoltages. In this case, we have $\chi \ll 1$, and link failures
reflect the anticipatory disconnection of lines to prevent damages of
the network and its components.  Other examples, besides electrical
power grids, are traffic systems, where overloaded streets cause
unreasonably long travel times along links, which may be interpreted
as effective link failures. The resulting choice of alternative routes
corresponds to a rebalancing of loads and is expected to cause
transient effects, with finite values of $\chi$.

As the model allows for effective simulations, it could also be useful
for close to real-time planning and optimization of network topologies
and load sharing, particularly for large networks.  Fully realistic
state-of-the-art simulation tools for, say, electrical power grids
that include network capacities, inductors, power generation, {\it
  etc.}, are computationally expensive and therefore not so well
suited for real-time simulation of large networks or their topological
optimization. Hence, simpler models could quickly and efficiently give
a useful overview that could serve as the starting point for more
detailed off-line simulations using classical power network
simulators.

\bigskip The authors gratefully acknowledge the support from the EU
Integrated Project IRRIIS (027568) and ESF COST Action P10 ``Physics
of Risk'' and comments by J.L. Mar\'{\i}n and A. Diu.

\bibliographystyle{apsrev}
\bibliography{paper2006-5}

\end{document}